# Photocatalytic oxidation in few-layer Tellurene for loss-invariant integrated photonic resonance trimming

*Dun Mao[1*], Yixiu Wang[1,2,3], Hwaseob Lee[1], Lorry Chang[1], Feifan Wang[1], Darren Wu[1], Yahui Xiao[1], Boshu Sun[1], Kaleem Ullah[1], Karl Booksh[4], Wenzhuo Wu[2], Tingyi Gu[1*]*

[1]Department of Electrical and Computer Engineering, University of Delaware, Newark, DE 19716
[2]School of Industrial Engineering, Purdue University, West Lafayette, IN 47907, USA
[3]Research Center for Humanoid Sensing, Zhejiang Lab, Hangzhou 3111100, China
[4]Department of Chemistry and Biochemistry, University of Delaware, Newark, Delaware 19716, USA

\* Email: dmao@udel.edu, tingyigu@udel.edu

Two-dimensional materials with unique physicochemical properties promote photocatalytic activities. As the 2D material composites research studies the statistical average of complex catalytic behaviors, an integrated photonic platform allows clean and single flake level photo-catalytic investigations with precisely quantified photocatalytic activities. In this paper, we track fluence-dependent photo-oxidation in two-dimensional Tellurene (2D Te) by the evanescently coupled micro-resonator. Nearly 32% of oxidation is achieved in ~10 nm 2D Te flake, compared to only 4.5% oxidation in 30 nm sample, probed by the resonance shift in silicon microring resonators (MRRs) substrate. The wider bandgap in the few layers of 2D Te allows faster charge transfer to adsorbed oxygen for a more efficient photocatalytic redox reaction. The photo-oxidation in hybrid 2D Te results in an invariant lineshapes of optical transmission resonance for wavelength trimming (more than 3× resonance bandwidth). The low threshold power, near-infrared, and in-waveguide resonance trimming scheme is compatible with most integrated photonic setups for easy fixing the nanofabrication-induced random resonance deviation for integrated photonic circuit applications in wavelength-division-multiplexing systems and spin qubits quantum computing

## 1. Introduction

The unique chiral chain atomic structure of two-dimensional tellurene (2D Te) [1-4] has exhibited intriguing electronic, thermal, and optical properties, [5-14] including its high carrier mobility (up to 700 cm$^2$V$^{-1}$s$^{-1}$ at room temperature), [2-4] layer-dependent bandgap (around 0.35 –1.6 eV), [7-8, 15] polarization selective infrared imaging, [9] exceptional second-order nonlinearities, [10-12] and unique photoinduced band splitting. [13] Compared to those photo-carrier excitation and light emission mechanics, 2D Te has less been explored for photochemistry, and photo-oxidation studies, which may guide to potential applications in catalysis, selective gas sensing, and photo energy harvesting. [16]

The research on the oxidation process in 2D materials has been focused on improving their ambient stability or tailoring the material's electronic properties. [17-18] The chiral chain atomic structure in 2D Te is stable and naturally protects the chalcogen from oxidation in ambient conditions. While the reports on baking-induced oxidation of trigonal Te (*t*-Te, a bulk form of chiral chain structure) or 2D Te are not found, laser-induced oxidation in *t*-Te has been observed by many groups. [19-22] Distinguished from thermal oxidation, above-bandgap photon excitation encourages the charge transfer in redox reaction and accelerates the atomic defects evolvement in 2D materials. A focused or waveguide-confined beam provides localized and deterministic control of the refractive index. [18] The micro-Raman spectroscopies revealed the laser-induced atomic structure transition pathways along *t*-Te, glass TeO$_2$, crystalline γ-TeO$_2$, β-TeO$_2$ to α-TeO$_2$. [20-22] The photo-oxidation experiments are only carried out in bulk Te with visible top excitation. The advancement of solution-grown 2D Te enables the study of such surface-dominant oxidation in van der Waals structures for device applications. Combined with the laser material processing with photo-oxidation and de-oxidation effects, it may enable the applications of low-power and highly efficient all-optical tuning devices, optical switches, and optical memories.

For a silicon ring resonator, nanofabrication variation (±0.5 nm) induced resonance deviation affects the applications of ring modulators, lasers, etc. Resonance trimming is needed for on-chip wavelength correction. As demanded by the system applications, [24-26] direct laser oxidation of the silicon layer has been the primary way of resonance trimming. [27-28] The activation energy of silicon photo-oxidation requires visible photons with three orders of magnitude higher fluence (532 nm, ~4J fluence) for reaching a similar wavelength trimming range (~0.2 nm) in photonic crystal cavities. Other volatile or nonvolatile on-chip resonance adjustment approaches are either associated with drastic loss variations, require sustaining power, or are subject to a small tuning range (Table S1 in Supporting Information file). To use another way for efficient resonance trimming with less loss and less energy consumption is essential. In this work, we report the evanescently coupled photon-oxidation in hybrid 2D Te with low-power in-waveguide excitation at telecommunication wavelength (0.8 eV photon energy).[29] Converting 2D Te ($n_{Te}$ = 4.7 at 1550 nm) to glass TeO$_2$ ($n_{g\text{-}TeO2}$ = 2.2 at 1550 nm) [23] leads to resonance shifts (comparable to a few resonance bandwidths) on a hybrid Te-covered silicon micro-resonator only by in-waveguide excitation of a few mJ. The hybrid 2D Te-Si waveguide has a few distinguished advantages (1) large resonance shift enabled by high refractive index contrast (53%) between 2D Te and Te oxide; (2) laser dosage independent collective optical loss ($\Delta k$<10$^{-5}$ or $\Delta n/\Delta k$>10$^3$); [30] (3) van der Waal (vdw) atomic structure protected material stability in ambient conditions; (4) balanced collective loss between material absorption and geometric scattering loss for loss independent resonance shift.

## 2. Results

The mechanism of the photocatalytic oxidation of 2D Te is illustrated in Figure 1a. In as prepared 2D Te (Figure 1a top inset), the bandgap ($E_g$) is inversely related to the thickness ($t$) or the number of layers ($n$): $E_g$ = 0.38 +1.8/$n$, where $n$ = $t$/3.91nm. [1-2, 6-7,14] The bandgap in a typical thin flake (10 nm) can reach 1.0 eV, compared to the bulk $E_g$ of 0.38eV. The wider bandgap of the thin flake leads to proper band offset for the photo-generated electrons to efficiently transfer from the conduction band to the adsorbed gas molecules, where the redox potential of $O_2/O_2$- (with an activation energy of 0.35eV [31]) is located within the bandgap (middle inset of Figure 1a). The formed superoxide anions $O_2$- are highly reactive and attack the Te-Te bond, resulting in Te-O bonds. [19-22] The oxidation may start from defect sites, and the formed random network of Te oxides is in the glassy state (Figure 1a bottom inset). [20-23] With increasing fluence/dosage, layer-by-layer oxidation is anticipated in the 2D Te. With increasing laser dosage, sequential formation of glass TeO$_x$, crystalline γ-TeO$_2$, β-TeO$_2$ to α-TeO$_2$ are anticipated, corresponding to previous experimental studies in powder form t-Te.[20-22] The penetration depth of the visible photon, as well as the oxide thickness, is approximately limited to 10 nm. [23]

The dynamics in the photocatalytic oxidation of 2D Te in ambient conditions are experimentally probed by near-infrared laser exposures. The 2D Te flakes were synthesized through a hydro-thermal synthesis approach. [2] Shutter-controlled 50 ns exposure of pulsed laser (center wavelength 1064 nm, average power 0.1 mW, repetition rate 600 kHz, pulse duration of 2 ns) was irradiated on 2D Te thin film. Generally, the sizes of Te flakes (>10μm) are much larger than the spot size of the Raman laser. During the laser exposure, the reflected image from the sample was monitored by the top CCD camera in situ (Supporting Information file II). The representative snapshots of the reflection image captured in the first 0.22s are selected from the recorded video (Figure 1b). The morphology evolvement of the sample during the exposure indicates the oxide starts to form within the first 10ms and stabilized at 0.22s. The final state was examined by the scanning electron microscope (bottom inset of Figure 1b) and atomic force microscope (inset of Figure 1c). For the 2D Te flake under test (~ 44 nm thick), reduced thickness near the center (15%) is associated with the Gaussian profile of the top incident beam and limited conductivity of the film. [32] The power-dependent dosage test was carried out with a 532 nm continuous wave (CW) laser (Supporting Information Section III, Figure S3). The intermediate micro-Raman spectra verify the emerging $TeO_x$ during the post-exposure (Figure S2-S3). Both visible and near-infrared laser exposures result in the formation of Te oxide with the diminishing Te portion (decreased $A_1$, $E_1$ peak intensity) at increasing laser fluence (Figure 1c).

Probed by micro-Raman, similar photo-oxidation is observed with in-waveguide excitations (Figure 2a). A cw TE polarized laser excitation (near 1539 nm, average power of 0.1 mW) is guided to interact with a thin flake of 2D Te by the single mode waveguide defined on a silicon-on-insulator substrate (Inset of Figure 2a). The mode is confined in the silicon waveguide with ~ 480 nm width and evanescently coupled to the 2D Te (Inset of Figure 2b). Confined by the waveguide, the converted area (480 nm) is smaller than the laser spot size of the micro-Raman spectrometer (~1µm). The micro-Raman spectrometer picks up signals from both the in-waveguide mode exposed area (orange curve in Figure 2a) and the area far away from the waveguide (blue curve in Figure 2a). After in-waveguide excitations, the micro-Raman spectra exhibit the new peak in lower frequency for the stable glassy $TeO_x$ (orange curve in Figure 2a). The evanescent wave is expended in z direction beyond 60 nm (blue curve in Figure 2b and Figure 2c). As the initial 2D Te thickness increases from 10 to 30 nm, the final oxide thickness reduces from 3.3 nm to 1.5 nm with similar exposure dosage (orange dots in Figure 2b, extracted from the MRR resonance shifts). It indicates that thinner films encourage photocatalytic oxidation, which might attribute to larger band offset facilitated charge transfer for the redox reaction (Figure 2d).

The power-dependent refractive index change in 2D Te was measured through the high $Q$ (~$10^5$) MRRs defined in the SOI substrate (Figure 3). Five MRRs with identical design (radius of 20 µm and MRR-bus waveguide gap near 200 nm) are side coupled onto the same bus waveguide (bottom right inset of Figure 3a). Due to fabrication variations, the MRRs with the same design came with slightly different resonance wavelengths (Figure 3a). Generated from a tunable laser, the cw light (<2 pm bandwidth) was launched into the MRRs through a fiber, polarization controller, a grating coupler, and bus waveguide. Light-matter interaction with 2D Te is triggered only when the wavelength of the input laser matches the resonance wavelength of the target MRRs. As each dip corresponds to one MRR, the dip marked in the orange arrow in Figure 3a exhibits a distinguished nonlinear response compared to the other four resonances. The insertion loss of each component is considered for deriving the effective power coupled into the MRR ($P_{in}$), which is estimated to be less than -10 dBm. The target MRRs highlighted in red in the bottom right inset of Figure 3a is shown in the optical microscope image in Figure 3b. Figure 3c shows a zoom-in view of the transmission spectra for the 2D Te-covered MRR. Resonance wavelength shifts around -70 pm as $P_{in}$ increases from -13 dBm to -10 dBm, which is close to the full width at half maximum (FWHM) of the resonance (Figure 3c). Figure 3d summarizes the transmission lineshapes evolution at increasing power levels, for both the monolithic control sample (marked in grey arrow in Figure 3a) and 2D Te-covered MRR. At such input power levels, no clear resonance shift is recorded in control MRR with a similar $Q$. The coupled mode theory (CMT) model was used to extract the exact intrinsic ($Q_{in}$), coupled ($Q_c$), and total quality factors ($1/Q_t = 1/Q_c + 1/Q_{in}$) from the nonlinear transmission spectra (dashed curves in Figure 3c). [29] $Q_{in}$ is sensitive to the change of extinction coefficient of the material, as well as nonlinear absorptions (such as two-photon absorption, free carrier absorption, etc). $Q_c$ is fixed by the geometry. Even with significant resonance shifts up to one FWHM, no additional/nonlinear absorption is reflected from those extracted $Q_{in}$, indicating the purely dispersive nonlinear response in 2D-Te undergoes structural transitions. The effective index of the single mode waveguide ($n_{eff\_Si}$=2.455), 10 nm 2D Te-covered waveguides ($n_{eff\_Te-Si}$=2.519), 11 nm glass $TeO_2$-covered waveguide ($n_{eff\_TeO2-Si}$=2.457) and confinement factor ($\Gamma$~2%) are calculated from the finite difference in time domain methods. The extracted effective index change ($\Delta n_{Te-TeO2}$ in Figure 3d and Figure 4d) after Te being oxidized to glassy $TeO_2$ is normalized to the covering length of Te flake (details can be seen in Supporting Information file section V). Note that crystalline $TeO_2$ has a slightly higher refractive index (~2.37) than the glass state (2.2), [23] which might reverse shift the resonance to a longer wavelength. However, this process is prohibited even at higher fluence, as the wide bandgap of $TeO_2$ (>1eV) prevents photo-absorption in telecommunication band and photothermal processes (the absolute bandgap of glass $TeO_2$ is not found in the literature but should be larger than $TeO_2$ polymorphs). [33]

To develop a more comprehensive understanding of the results, we repeated the experiments on 2D Te with different thicknesses. The solution synthesized 2D-Te flakes were firstly drop cast onto a PDMS stamp. Then, a flake with the proper size, quality, and thickness is chosen and transferred onto a selected set of MRRs with similar $Q_s$ (~$10^5$) through stamp-transferring processing on a hot plate (details provided in Supporting Information file IV). Figure 4 shows the nonlinear measurement result for a thicker 2D-Te (~ 30 nm measured by AFM) and longer coverage length $L$ on MRR (~ 15 µm). The collective nonlinear transmission spectra are shown in Figure 4a. Like the results of MRR covered by thinner and smaller Te flakes in Figure 3, one MRR exhibits distinguished blue-shifting resonance at increasing $P_{in}$ (from -15 to -9 dBm). The optical image of the sample under test is given in Figure 4b. With longer coverage length and similar excitation power level, resonance shifts up to -290 pm are observed in the MRR with FWHM of 120 pm (Figure 4c-d). The CMT extracted $Q_{in}$ of the hybrid MRR keeps constant at increasing power levels along with the significant resonance shift, which might be attributed to the balance between reduced material absorption and rising scattering loss.

We summarized the linear and nonlinear responses in MRRs in Figure 5. The oxidation-induced resonance trimming of the hybrid MRR with 2D Te is summarized in Figure 5a. The resonance shift is normalized with the coverage length. Under the same input fluence, a larger effective index change is observed for the thinner sample (10 nm 2D Te), normalized by the coverage length ($L$). No oxidation/permanent effective index shift is observed in thick samples (t > 50nm). In the 30nm thick sample, the oxidation percentage saturates at the input fluence of more than 2.0 mJ. Further and more energy-efficient oxidation is observed in the thinnest sample (t = 10nm), which is consistent with the hypothesis in Figure 2c. The thinner sample would have higher trimming efficiency than the thick one. Figure 5b compares the transmission spectrum of one MRR before (blue) and the transmission spectrum of the same MRR after transferring the 2D Te (orange). The dots are experimental data, and the dashed curves are CMT fits. It is noted that the pre-and post-transferring results in this comparison are obtained from the same device, to exclude the nanofabrication-induced scattering loss variations among the devices. For this device under test, the intrinsic quality factor $Q_{in}$ dropped from 71k to 46k with 2D Te so that the resonance dip becomes less sharp due to the material absorption and scattering loss. From the CMT extracted $Q_{in}$, the total intrinsic loss rate (related to material absorption and scattering loss) in MRR can be derived $\gamma_{in}=\omega_0/Q_{in}$, where $\omega_0$ is the angular frequency of the resonance. The extrinsic loss from the 2D Te absorption and scattering was calculated as the difference between the intrinsic loss rate of the monolithic MRR before ($\gamma_{in\_o}$) and after transferring ($\gamma_{in\_Te}$): $\Delta\gamma_{in} = \gamma_{in\_Te} - \gamma_{in\_o}$. The results of three samples with different coverage lengths $L$ are summarized in the inset of Figure 5b (details of fitting results for the other two devices can be seen in Supporting Information file VI). Sublinear relationship between intrinsic loss rate ($\Delta\gamma_{in}$) and $L$, varied by the flake thickness. The radius of MRR for the three samples is kept the same as 20 µm.

## 3. Conclusion

The reduced layer thickness in 2D Te results in wider bandgap, higher band-offset and more efficient charge transfer for enhancing the photocatalytic reaction rate. The transition of t-Te to glass Te-oxide has been explored in powder form, but the layered dependent response is reported first time here. This observation is enabled by (1) solution synthesize thin film flake and (2) tracking the effective index change through the guided wave confined in the waveguide substrate. The high Q resonator allows sensitive tracking of refractive index change down to $10^{-3}$, providing a quantitative characterization of the photocatalytic product. More efficient and throughout oxidation is probed in thinner 2D Te through the evanescently coupled field, which might be attributed to the faster charge transfer for redox reactions. This photocatalytic response in 2D materials have been investigated for water splitting or other clean energy applications in a large quantity. [16, 34-35] Here we focus on single flake level study and revealed alternative potential applications in nanophotonic devices. The on-demand resonance trimming by in-waveguide photo-oxidation is prominent and permanent, for back-end-of-line compatible resonance trimming. The sustaining power-free and shape-preserved resonance correction are critical for system applications involving multi-wavelength operations, such as low-power optical interconnects in wavelength division multiplexing systems and spin qubits quantum computing in silicon photonics. [24-28]

## 4. Experimental Methods

*Photonic chip post-processing process*: The foundry-fabricated low-loss photonic micro-ring resonators are based on a silicon-on-insulator (SOI) substrate with an oxide cladding layer. The thickness of the top cladding layer, the silicon layer, and the oxide isolation layer are 5.4 µm, 220 nm, and 2 µm, respectively. The top oxide cladding layer is removed by etching in the buffered oxide etchant 1:6, details can be seen in Supporting Information file IV.

*Material transfer process*: The hydro-thermal synthesized Te flakes can be transferred onto silicon chips through dry and wet transfer processes. Details are shown in Supporting Information file IV.

*Optical measurements*: cw light generated from a tunable laser was coupled into the chip through a polarization controller and single-mode fiber and coupled onto the hybrid waveguide through a grating coupler. The output light was collected by a power meter.

**Supporting Information**

Supporting Information is available from the Wiley Online Library or from the author.

**Acknowledgments**

The authors greatly appreciated Prof. Matthew Doty from the Department of Materials Science and Engineering at the University of Delaware for valuable discussions. This work was supported by Army Research Office W911NF-20-1-0078.

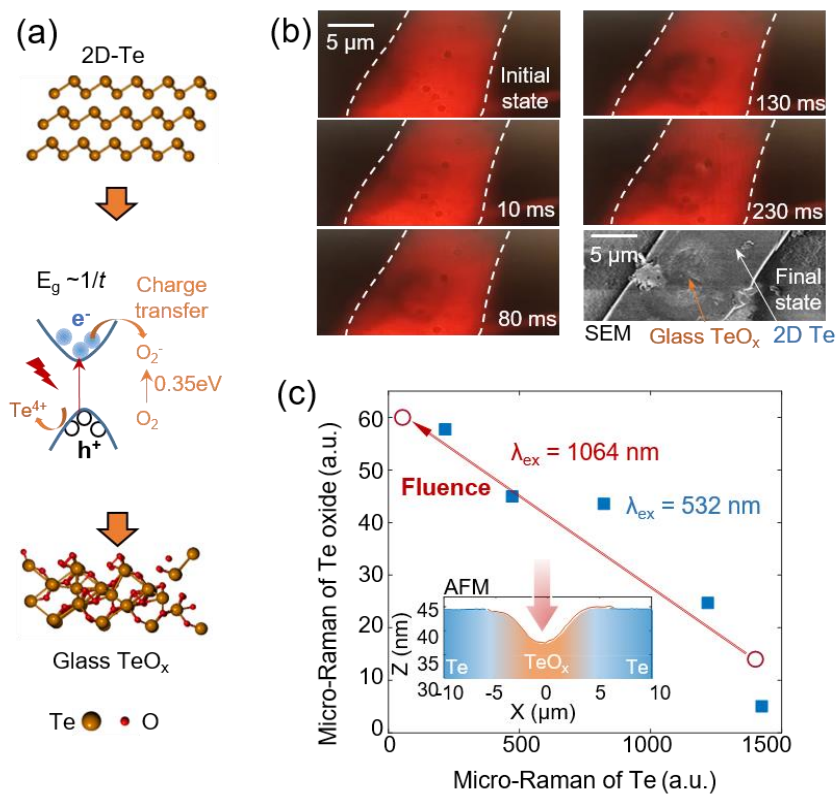

**Figure 1. Photocatalytic oxidation in layered Te by top laser irradiation**. a) Physiosorbed $O_2$ - dissociating and inserting into the photo-activated Te-Te bond. Top inset: ball and stick pictures of initial 2D Te; Middle inset: the diagram for photoreaction; Bottom inset: the formed glass $TeO_2$. b) In-situ monitoring of the photo-oxidation dynamics of the 2D-Te under 1064 nm nanosecond laser top excitation (~45nm thick). The scanning electron microscope (SEM) (bottom inset) examines the final state morphology. c) Correlation between Micro-Raman intensity of formed glass $TeO_2$ (at low frequency < 90 cm$^{-1}$) and $A_1$ peak of 2D Te, with increasing fluence from top incident near-infrared (orange, Figure S2) and visible laser (blue, Figure S3). Inset: atomic-force microscope measured topology of the final state in b.

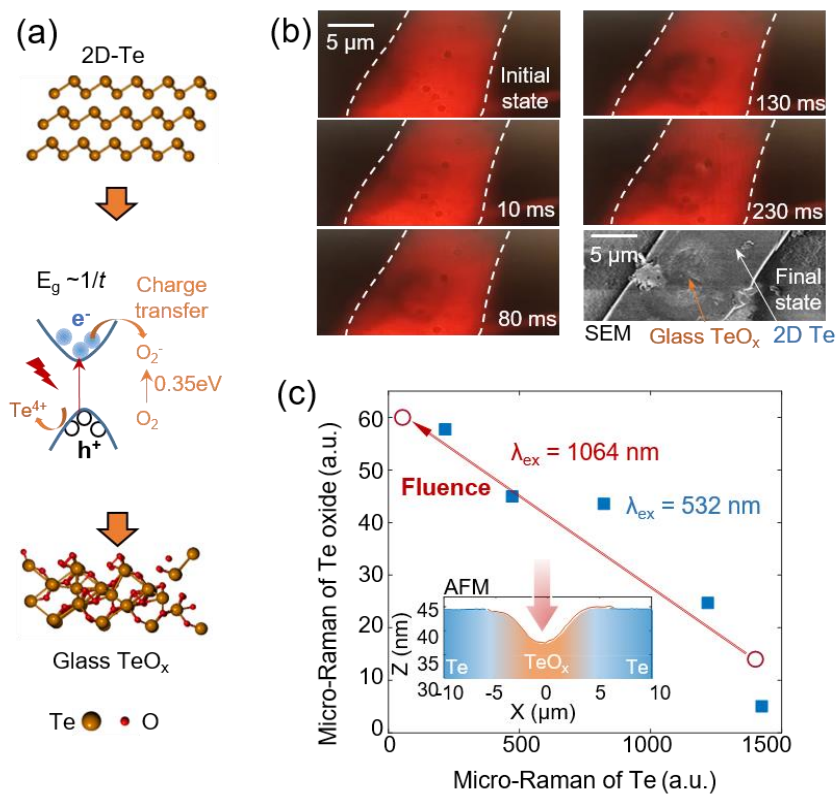

**Figure 1. Photocatalytic oxidation in layered Te by top laser irradiation**. a) Physiosorbed $O_2$ - dissociating and inserting into the photo-activated Te-Te bond. Top inset: ball and stick pictures of initial 2D Te; Middle inset: the diagram for photoreaction; Bottom inset: the formed glass $TeO_2$. b) In-situ monitoring of the photo-oxidation dynamics of the 2D-Te under 1064 nm nanosecond laser top excitation (~45nm thick). The scanning electron microscope (SEM) (bottom inset) examines the final state morphology. c) Correlation between Micro-Raman intensity of formed glass $TeO_2$ (at low frequency < 90 cm$^{-1}$) and $A_1$ peak of 2D Te, with increasing fluence from top incident near-infrared (orange, Figure S2) and visible laser (blue, Figure S3). Inset: atomic-force microscope measured topology of the final state in b.

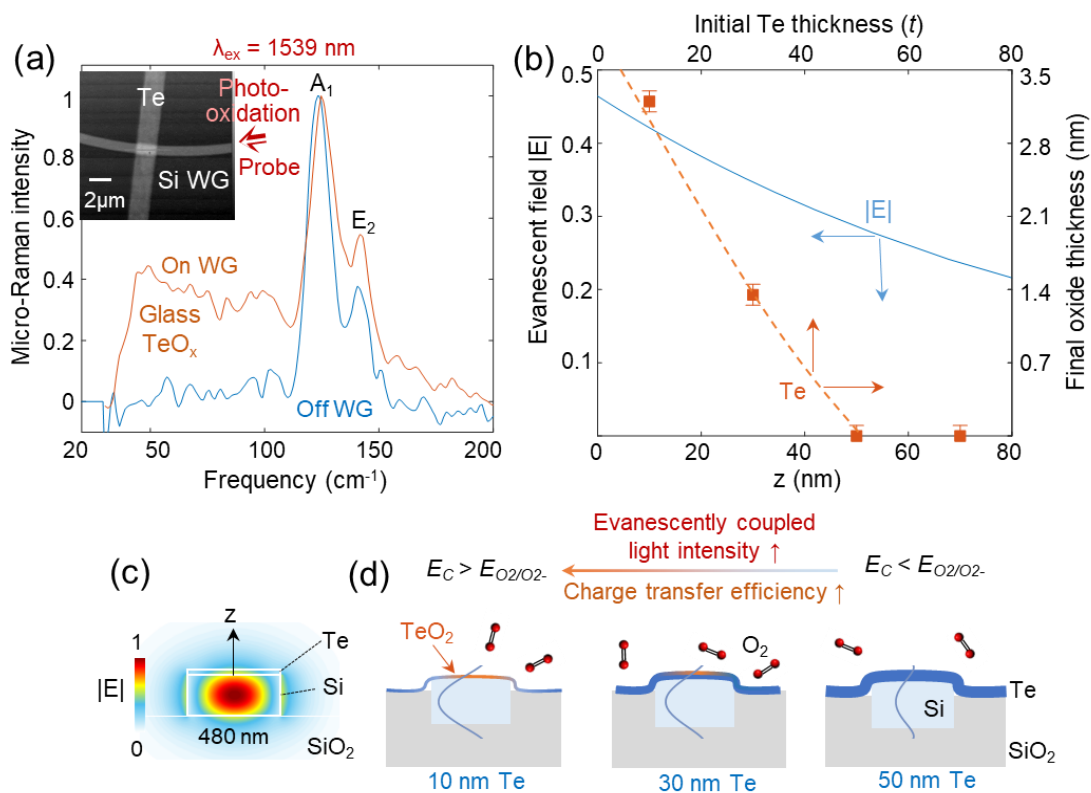

**Figure 2. Evanescently coupled wave for photo-oxidation and in-situ probing of oxide layer thickness in layered Te.** a) Micro-Raman spectra of 2D Te on (orange) and off (blue) silicon waveguide, after substrate guided wave exposure of 1550 nm cw laser (SEM in top left inset). b) Calculated evanescent field intensity versus z-distance to waveguide top surface (blue), and the experimentally measured final oxide thickness at different the initial 2D-Te thickness (measured by AFM, orange), extracted from the measured resonance drift as specified in Fig. 3-4. (c) Simulated mode profile on the cross section of the single mode silicon wave (SiWG in the inset of a). The waveguide width and thickness are 480 nm 220 nm, respectively. (d) Mechanism of the in-waveguide excitation-induced oxidation. A thinner film with a wider bandgap and higher conduction band ($E_c$) results in more efficient charge transfer to physically adsorbed $O_2$. Red curve: simulated E field along z direction superimposed on the device geometry.

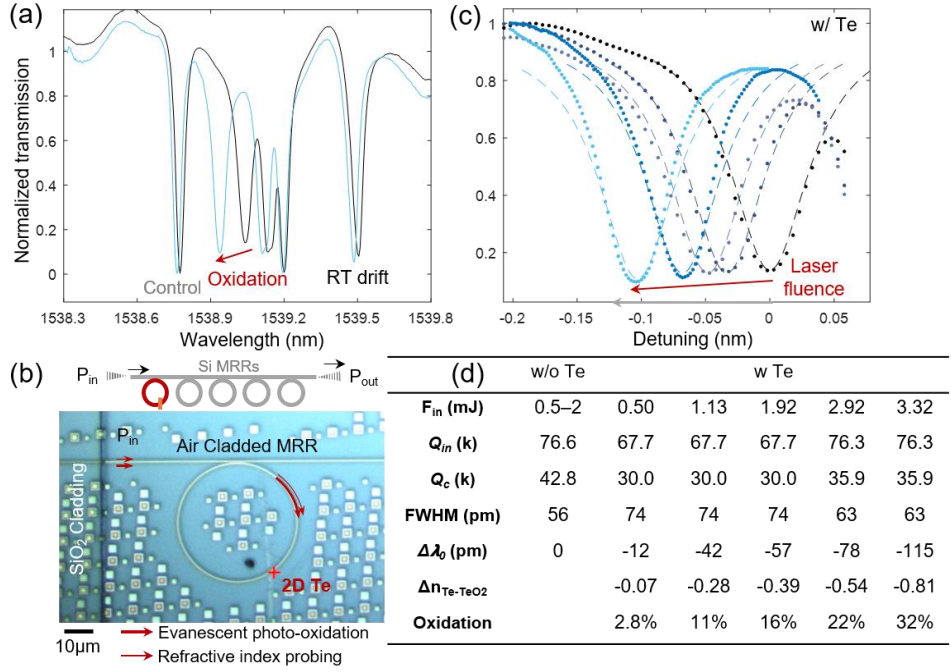

**Figure 3. Photo-oxidation in 10 nm thick 2D Te on silicon microring resonator.** a) Measured transmission spectra of a set of MRRs coupled to the same bus waveguides. The MRR covered by Te exhibits distinguished nonlinear resonance shifts after in-waveguide exposure, compared to the other four MRRs (marked in red). b) Optical microscope image of the hybrid MRR covered by 2D Te (10 nm thick and 1.6 µm long). Top inset: photonic circuits schematics. c) Transmission spectra of the MRR in b. Resonance blue shift is observed with higher in-waveguide excitations. Circles: experimental data. Dashed curves: correspondent coupled mode theory (CMT) fitting. d) Summarization of extracted parameters from the hybrid MRR and one with similar extinction ratio and quality factor (indicated by the grey arrow in a). Significant resonance shift ($\Delta\lambda$) is observed with the micro-Raman probed structural transitions (Figure 2a), without noticeable change of intrinsic ($Q_{in}$), coupling ($Q_c$), and total quality factors ($Q_t$). The refractive index change ($\Delta n_{Te-TeO2}$) and the Te film's oxidation percentage are estimated from the hybrid waveguide's effective index change.

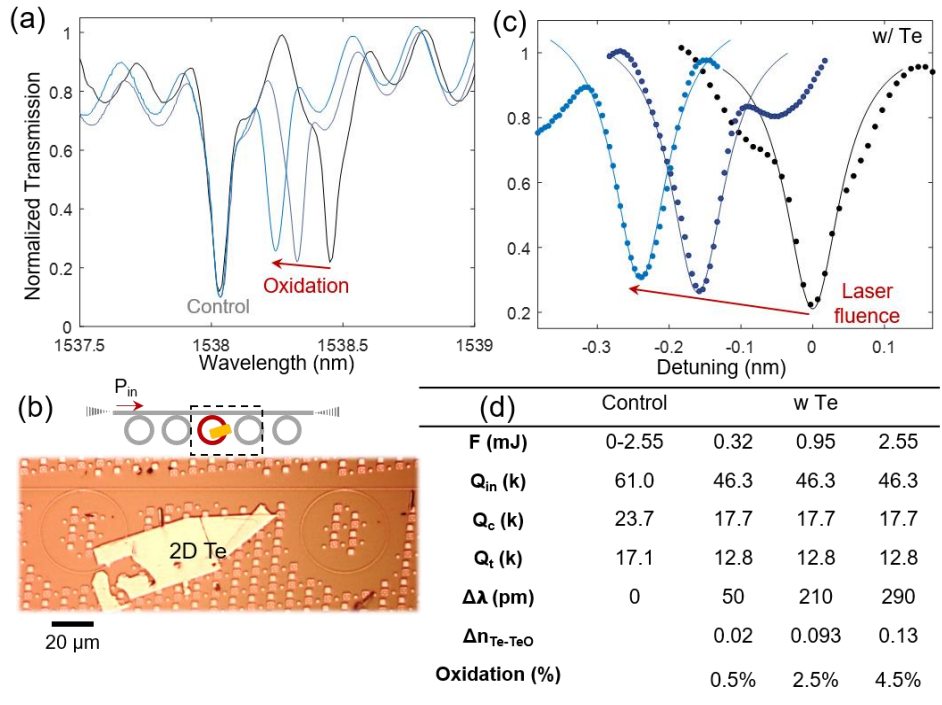

| (d) | Control | | w Te | | |
|---|---|---|---|---|---|
| F (mJ) | 0-2.55 | 0.32 | 0.95 | 2.55 |
| $Q_{in}$ (k) | 61.0 | 46.3 | 46.3 | 46.3 |
| $Q_c$ (k) | 23.7 | 17.7 | 17.7 | 17.7 |
| $Q_t$ (k) | 17.1 | 12.8 | 12.8 | 12.8 |
| $\Delta\lambda$ (pm) | 0 | 50 | 210 | 290 |
| $\Delta n_{Te-TeO}$ | | 0.02 | 0.093 | 0.13 |
| Oxidation (%) | | 0.5% | 2.5% | 4.5% |

**Figure 4. Photo-oxidation in 30 nm thick 2D Te hybrid silicon MRR**. a) Measured transmission spectra of a 2D Te at different laser exposure fluence. b) Schematics of the device settings and the optical image of the region in the dashed square. Bottom: optical microscope image of the device under test (Te thickness ~30 nm, coverage length ~ 15 µm). c) Zoom-in transmission spectra of the MRR with Te, with input powers set from -15 to -9 dBm. Circles: experimental data. Dashed curves: CMT fitting. d) Summarization of extracted parameters from the measured transmission spectra from hybrid MRR and the control resonance (identified in a).

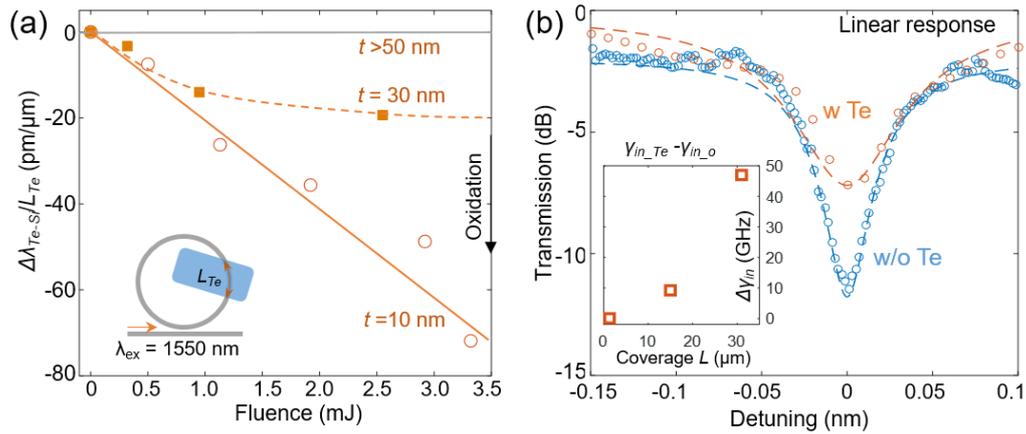

**Figure 5. Waveguide excitation-driven oxidation of hybrid integrated 2D Te and its layer thickness dependence**. a) Normalized resonance shift by the coverage $L$ (indicated in the inset) versus the fluence of the photon coupled into the MRR, for 10 nm thick flake (empty circles), 30 nm flake (solid squares), and thicker films (grey). Lines and curves are eye guides. b) Comparison of the same MRR before (blue) and after transferring Te (orange curve). Inset: Incremental total intrinsic loss ($\gamma_{in\_Te} - \gamma_{in\_o}$) introduced by 2D Te with the coverage length ($L$) along the MRRs. Circles: experimental data. Dashed curves: CMT fitting.

Supporting Information

**Photocatalytic oxidation in few-layer Tellurene for loss-invariant integrated photonic resonance trimming**

*Dun Mao[1*], Yixiu Wang[1,2,3], Hwaseob Lee[1], Lorry Chang[1], Feifan Wang[1], Darren Wu[1], Yahui Xiao[1], Boshu Sun[1], Kaleem Ullah[1], Karl Booksh[4], Wenzhuo Wu[2], Tingyi Gu[1*]*

**I. Comparison to other resonance trimming schemes in silicon micro-resonators**

**Table S1**. Comparison of resonant wavelength tuning schemes in silicon micro-resonators

| Mechanism | Additional source? | Laser wavelength | Fluence | Hybrid WG $\Delta n_{eff}$ | Quality factor change |
|---|---|---|---|---|---|
| Silicon photo-oxidation [27-28] | Top exposure | 532 nm | 120J (cw) | ~0.01 (derived) | Small (-62% [28], 54% [27]) |
| Hybrid 2D Te photo-oxidation (**this work**) | No | 1550 nm | 2.55 mJ (cw) | 0.062 [b] | No |
| Hybrid GST PT [a] [S1-S2] | No | 1550 nm | Sub-nJ (pulsed) | 0.06 +0.03i [b] | Large contrast |
| Hybrid GSST PT [S3] | Top exposure | 633 nm +780 nm | 13-400 nJ (pulsed) | 0.04 +0.007i [b] | Small |
| Hybrid vdW $In_2Se_3$ PT [S4] | Top exposure | 1064 nm | 0.25 nJ (pulsed) | 0.03 [b] | Small |
| Silicon strain [c] [S5] | Strain/stress | -- | -- | $10^{-4}$ | Small |
| Thermal tuning [c] | Integrated heater | -- | -- | $10^{-2}$ | No |
| Carrier injection [c] [S6] | Integrated p/n junction | -- | -- | $10^{-4}$ | ↓(absorption) |

[a] PT: Phase transition; [b]: The effective index change of the hybrid waveguide is estimated based on 10 nm thick film for Te, GST, GSST, and 50 nm layered $In_2Se_3$. [c] Grids filled with grey: approaches used in monolithic silicon photonics but requires sustaining excitation power and additional substrate cooling

**II. House-made setup for 1064nm laser top exposure and in-situ morphology monitoring**

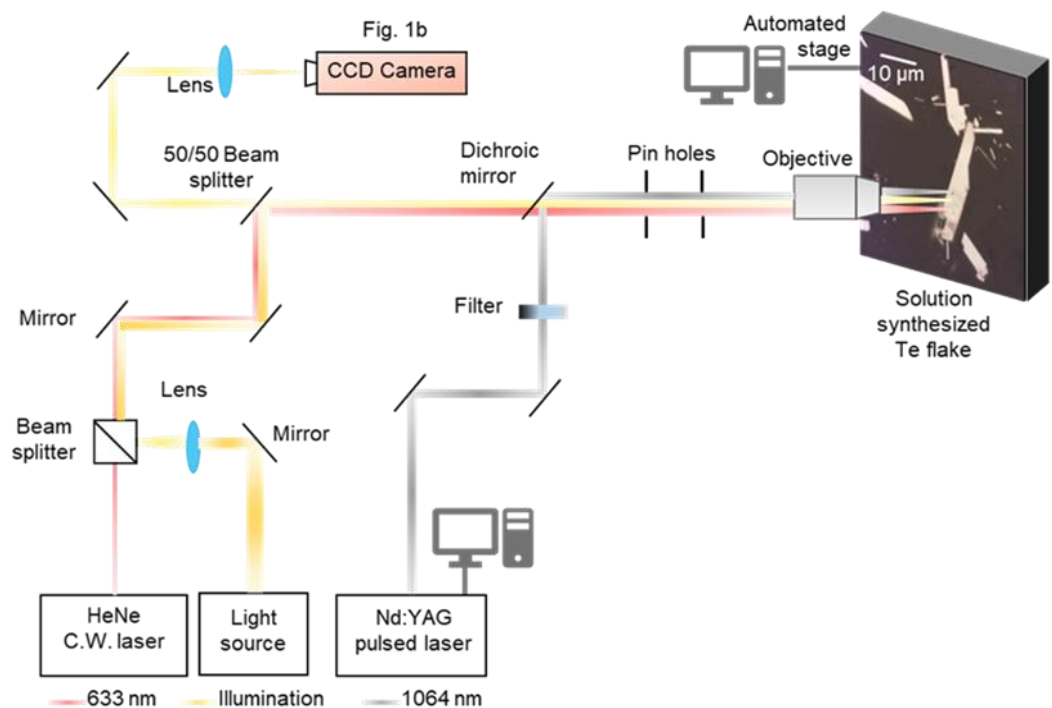

**Figure S1. Optical setup for 1064 nm laser processing of solution synthesized 2D Te flakes**. Schematics of direct laser writing setup. Three light sources are coupled onto the chip for alignment (633 nm cw laser), illumination (Tungsten bulb), and laser writing (1064 nm, ~ 1ns pulse duration, and 600 kHz repetition rate). The sample is mounted on a piezo-controlled xyz linear translation stage. [S7]

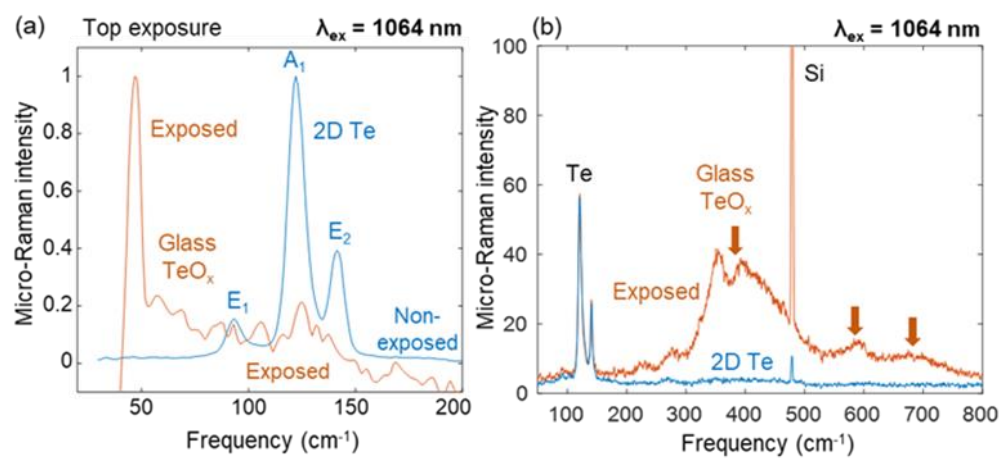

**Figure S2. MicroRaman probed oxidation in 2D Te with 1064nm pulsed laser top excitations**. a) On- (orange) and off- (blue) exposed spot. b) Broadband spectra exhibiting $TeO_x$ peaks.

## III. Top illumination processing with 532 nm cw laser

At increasing fluence (controlled by laser power with the same exposure time), the broadband background signal coming from glassy Te oxide rises with the laser power. An opposite trend is observed for the $A_1$ and $E_1$ peaks for 2D Te. Note that the in-situ micro-Raman spectra are collected during the high-power exposure (10% -100%), and the post-irradiation spectra are collected with 1% power to eliminate nonlinear material response.

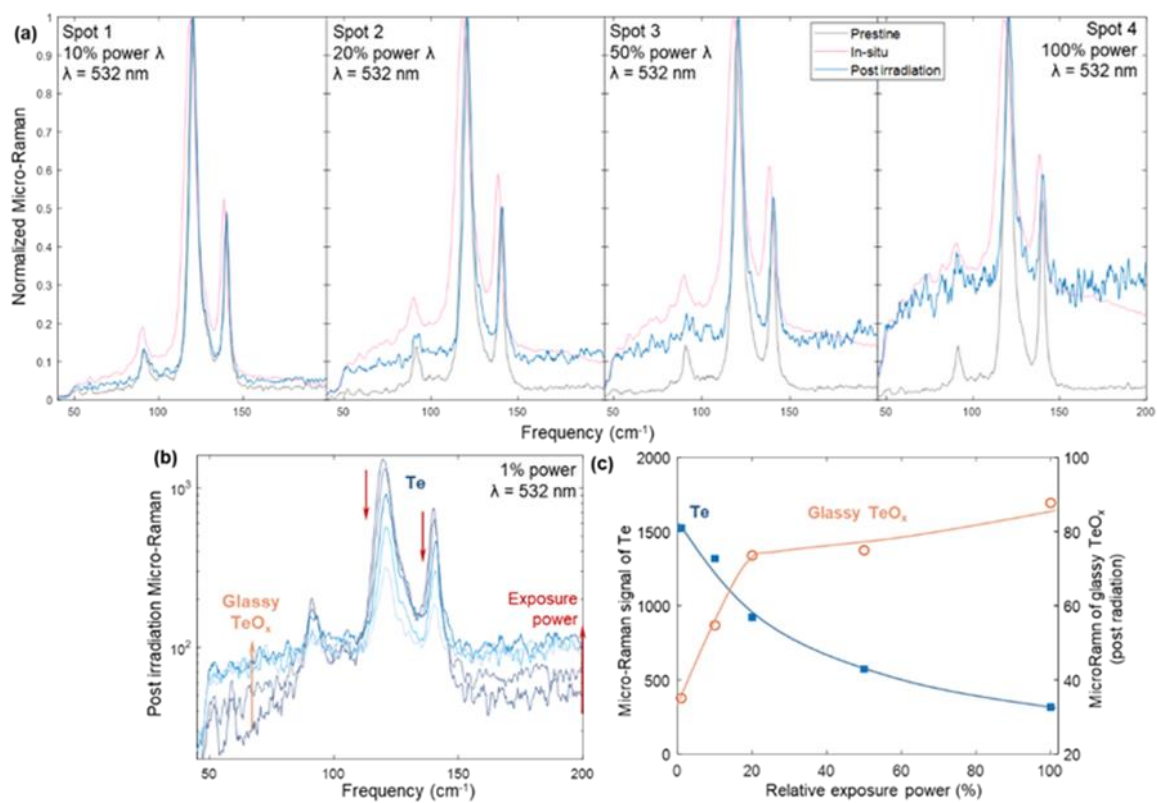

**Figure S3. Micro-Raman probed oxidation in ~20 nm 2D Te with 532 nm c.w. laser top excitations**. a) Normalized micro-Raman spectra of four different spots across flakes on the pristine sample (grey), in-situ monitoring at different exposure laser power (pink) and post-exposure (blue). b) Spectra of fluence-dependent post-exposure sample probed with low power. c) Micro-Raman signal level from Te oxide (~60 cm$^{-1}$ marked in b) with orange arrows and Te ($A_1$ peak) versus the laser power (labeled in a)).

## IV. Schematics of 2D Te integration on silicon photonics

Firstly, we prepare the high-quality silicon photonic substrate manufactured in the foundry.[S8] The top oxide is removed to expose the silicon microring resonator (MRR), and the rest of the supporting waveguide structures and couplers are protected under the thick oxide (Figure S4a-c). The solution-grown 2D Te are integrated into the arrays of high-quality MRRs (schematics, Figure S4d) through wet transferring (Figure S4ef) and dry transferring (Figure S4gh). To obtain high-quality 2D Te flake covered on the desired position of silicon devices, we utilize this dry transfer method.

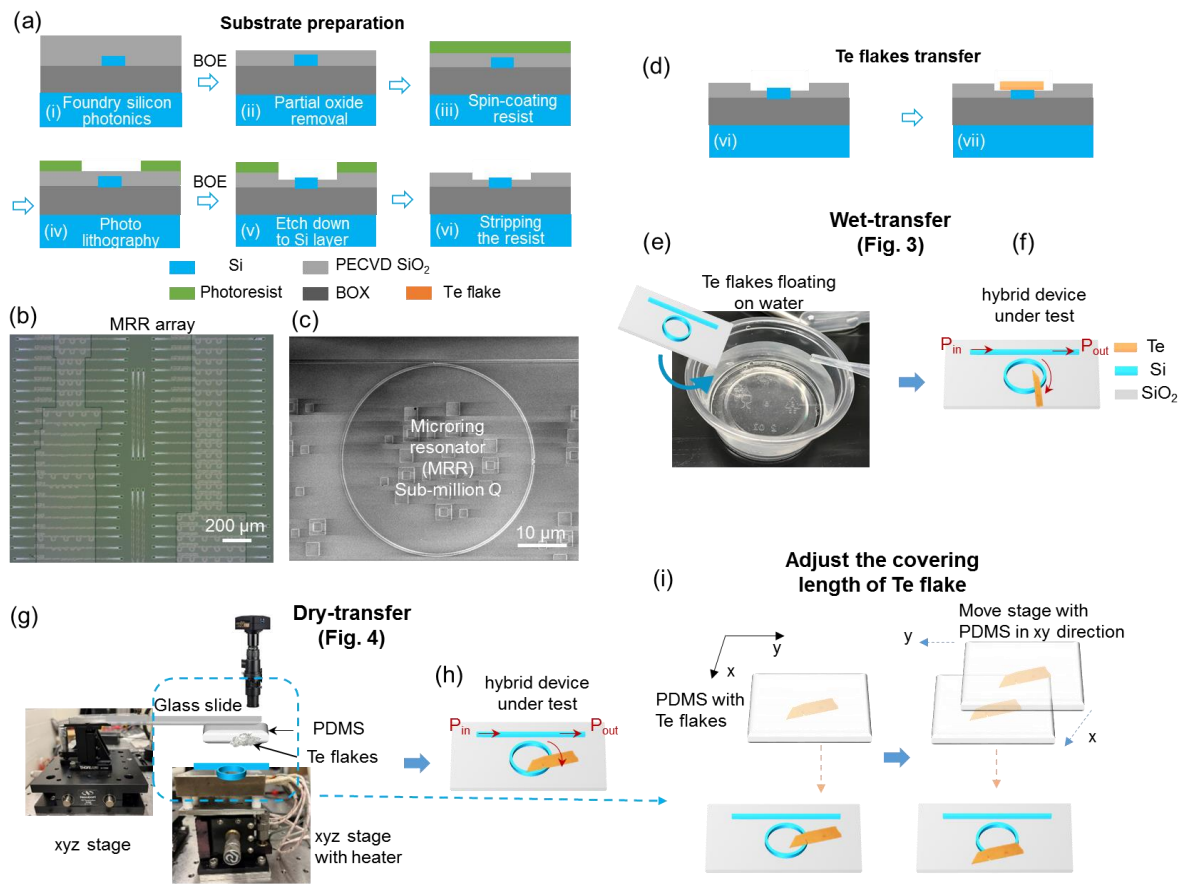

**Figure S4. Schematics of device preparation**. a) post-processing steps of the foundry-manufactured high-quality silicon MRRs. b) Optical microscope image of the open window after stripping the resist. c) Zoom-in SEM image of the exposed MRR to verify the exposure of the silicon device layer. d) Schematics of Te flakes transfer process. e) Wet transfer and g) dry transfer process and correspondent devices (f, h). We only show two specific devices which results are fully analyzed in the main text. i) Schematics of adjusting the covering length of Te flakes in the dry transfer process.

Take one droplet of tellurium raw solutions to drop on one synthesized transparent PDMS on a glass slide. After solvent evaporation (~5 minutes), we flipped the glass with the PDMS side facing the SOI substrate and mount the glass slide on a separate xyz precision positioner (Figure S4g). The open-window silicon photonic sample is mounted on the other XYZ stage with a heating function. After aligning the desired 2D Te flake and the silicon MRR under the top camera, the glass slide landed on the silicon photonic substrate. Then we turn on the heating stage to bring the sample temperature to 70℃ for 10 mins, to release the Te flake onto the target MRR. After releasing Te flakes, the PDMS-on-glass substrate is detached and removed (Figure S4h). For controlling the covering length of Te flakes over the silicon ring resonator, it can be done by moving the stage with PDMS in xy direction (Figure S4i). By carefully moving and aligning PDMS with the Te flake to the desired position on the silicon ring resonator with controllable covering length, it can be landed and released the Te flake on the silicon substrate.

**V. Effective index extraction from the transmission spectra**

By curve fitting the transmission spectra in Figure 3c and Figure 4c, the effective index change can be extracted after Te oxidized to glassy TeO$_2$. The transmission spectrum is shown as: [S9]

$$T = \frac{|\gamma - a*e^{-j\theta}|^2}{|1 - a*\gamma*e^{-j\theta}|^2}, \tag{S1}$$

where $\gamma$ is the transmission factor of a ring resonator, $\alpha$ is the linear loss of a ring, the phase $\theta = \frac{2\pi}{\lambda}n_{eff}L$, and $L$ is the effective length.

By fitting the measured transmission spectra to the equation above, it is able to get the parameters of $\gamma$, $\alpha$, and phase $\theta$.

In this case of the hybrid device, the phase is contributed both from the Si waveguide without Te covering and Te covered Si waveguide: [S9]

$$\theta = \frac{2\pi}{\lambda}n_{eff\_Si}L + \frac{2\pi}{\lambda}n_{eff_{Te-Si}}L_{Te}, \tag{S2}$$

where $L_{Te}$ is the covering length of the Te flake ($L_{Te}$ can be estimated from the optical image of the hybrid device), $L$ is the perimeter of the Si ring ($L=2\pi R$, $R=20$ μm), $n_{eff\_Si}$ is the effective index of Silicon waveguide, $n_{eff\_Te-Si}$ is the effective index of Te covered Si waveguide.

After Te is oxidized to glassy-TeO$_2$, the phase difference (Δθ) is only attributed to the effective index change due to oxidation:

$$\theta + \Delta\theta = \frac{2\pi}{\lambda}n_{eff\_Si}L + \frac{2\pi}{\lambda}(n_{eff_{Te-Si}} + \Delta n_{eff_{Te-TeO2}})L_{Te}, \tag{S3}$$

where $\Delta n_{eff\_Te\text{-}TeO2}$ is the effective index change (corresponding to the values in Figure 3d, Figure 4d) due to Te oxidized to glassy TeO$_2$. The extraction of effective index change ($\Delta n_{eff\_Te\text{-}TeO2}$) due to oxidation has considered the effect of Te flake length ($L_{Te}$) from the equations above.

**VI. 2D Te covering length-dependent intrinsic loss in silicon microring resonator**

Other two hybrid Te-silicon devices have also been characterized to see how the intrinsic Q factor changes after being transferred with one Te flake. The hybrid device in Figure S5a has a thickness of 20 nm and a covering length of 1.3 µm. The intrinsic Q factor dropped a little from 152.2k to 151.9k (Figure S5b). Another hybrid device in Figure S4c has a thickness of over 30 nm and a covering length of 31 µm. The intrinsic quality factor dropped from 29.8k to 13.9k (Figure S5d).

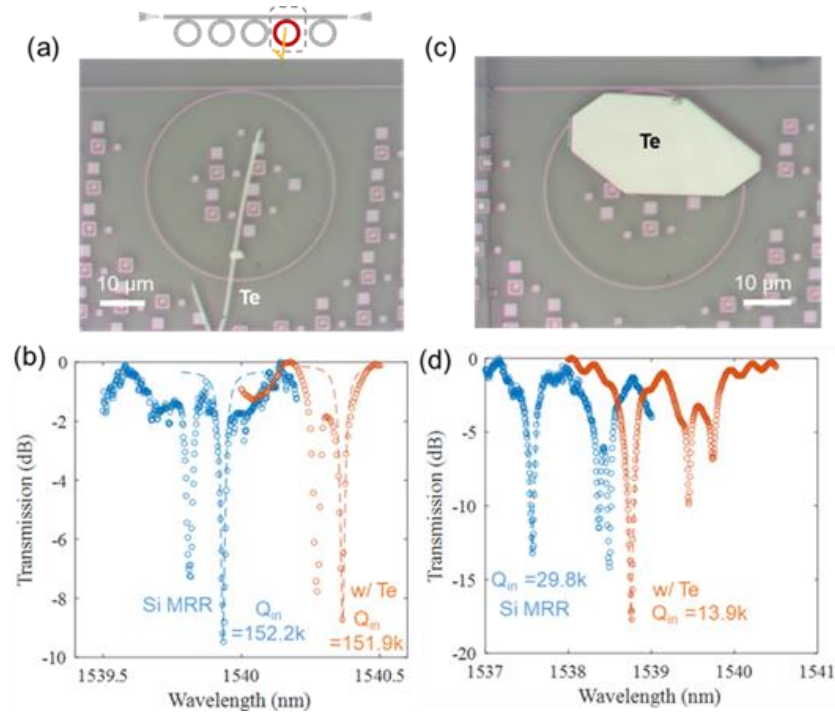

**Figure S5. Exemplary devices showing the Te covering length-dependent intrinsic loss in silicon microring resonator (MRR)**. a) The optical microscope image of an MRR covered with ~2µm covering the length of transferred Te flake. b) Corresponding transmission spectra in the MRR before transferring (with top oxide removed) (blue) and the same device after transferring (red). The resonance shift is attributed to the environmental temperature fluctuation over days. c) Another device with a much longer covering length (31 µm). d) Transmission spectra of the device in c before (blue) and after (red) transferring the Te flake.